\documentclass[12pt,a4paper,twoside,english]{article}
\usepackage[LGR,T1]{fontenc}
\usepackage[latin1]{inputenc}
\usepackage{amssymb}

\makeatletter

\DeclareRobustCommand{\greektext}{%
  \fontencoding{LGR}\selectfont\def\encodingdefault{LGR}}
\DeclareRobustCommand{\textgreek}[1]{\leavevmode{\greektext #1}}
\DeclareFontEncoding{LGR}{}{}
\DeclareTextSymbol{\~}{LGR}{126}

\newcommand{\lyxaddress}[1]{
\par {\raggedright #1
\vspace{1.4em}
\noindent\par}
}

\makeatother

\usepackage{babel}
\begin{document}

\title{Coupling of (ultra-) relativistic atomic nuclei with photons}

\author{\noindent {\normalsize{M. Apostol$^{1,a}$ and M. Ganciu$^{2,b}$}}}

\maketitle

\lyxaddress{\noindent $^{1}$Institute of Atomic Physics, Institute for Physics
and Nuclear Engineering, Magurele-Bucharest 077125, MG-6, POBox MG-35,
Romania,$^{2}$National Institute for Lasers, Plasma and Radiation
Physics, Magurele-Bucharest 077125, POBox MG-36, Romania }
\begin{abstract}
The coupling of photons with (ultra-) relativistic atomic nuclei is
presented in two particular circumstances: very high electromagnetic
fields and very short photon pulses. We consider a typical situation
where the (bare) nuclei (fully stripped of electrons) are accelerated
to energies $\simeq1TeV$ per nucleon (according to the state of the
art at LHC, for instance) and photon sources like petawatt lasers
$\simeq1eV$-radiation (envisaged by ELI-NP project, for instance),
or free-electron laser $\simeq10keV$-radiation, or synchrotron sources,
etc. In these circumstances the nuclear scale energy can be attained,
with very high field intensities. In particular, we analyze the nuclear
transitions induced by the radiation, including both one- and two-photon
proceses, as well as the polarization-driven transitions which may
lead to giant dipole resonances. The nuclear (electrical) polarization
concept is introduced. It is shown that the perturbation theory for
photo-nuclear reactions is applicable, although the field intensity
is high, since the corresponding interaction energy is low and the
interaction time (pulse duration) is short. It is also shown that
the description of the giant nuclear dipole resonance requires the
dynamics of the nuclear electrical polarization degrees of freedom. 


\emph{}\\

\noindent \emph{$^{a}$}Electronic mail: apoma@theory.nipne.ro

\noindent $^{b}${\normalsize{Electronic mail: mihai.ganciu@inflpr.ro }}{\normalsize \par}
\end{abstract}

\section{\noindent Introduction. Accelerated ions }

It is well known that the nuclear photoreactions occurr in the $keV-MeV$-energy
range. In particular, the characteristic energy of the giant dipole
resonance (which implies oscillations of protons with respect to neutrons)
is $10-20MeV$.$^{1-4}$ In order to get this energy scale typical
mechanisms are used, like Compton backscattering (for instance a laser-electron
system), or electron bremsstrahlung (usually with the same nucleus
acting both as converter and target), etc.$^{5-18}$ High intensity
laser pulses can be used for accelerating electrons in compact laser-plasma
configurations.$^{3,17}$ High-power and short-pulsed lasers are pursued
nowadays for increasing the intensity of the electromagnetic field.$^{19}$
Photon-ion or photon-photon mediated ion-ion interactions are also
well known in the so-called peripheral reactions.$^{20,21}$ Vacuum
polarization effects have also been discussed recently in high-energy
photon-proton collisions,$^{22}$ or light-by-light scattering in
multi-photon Compton effect.$^{23-25}$ We describe here a high-energy
and high-field intensity coupling of the atomic nucleus to photons
from various sources (\emph{e.g.}, optical laser, free electron laser,
synchrotron radiation) by using (ultra-) relativistic atomic nuclei. 

We consider (ultra-) relativistically acelerated ions moving with
velocity $v$ along the $x$-axis. We envisage acceleration energies
of the order $\varepsilon=1TeV$ per nucleon (according to the state
of the art at LHC, for instance).$^{26}$ At these energies the ion
is fully stripped of its electrons, so we have a bare atomic nucleus.
We assume that a beam of photons of frequency $\omega_{0}$ is propagating
counterwise (from a laser source, or a free electron laser, or a synchrotron
source, etc), such that the photons suffer a head-on collision with
the nucleus. The moving nucleus will \textquotedbl{}see\textquotedbl{}
a photon frequency 
\begin{equation}
\omega=\omega_{0}\sqrt{\frac{1+\beta}{1-\beta}}\,\,,\,\,\beta=v/c\label{1}
\end{equation}
 in its rest frame, according to the Doppler effect. For (ultra-)
relativistic nuclei ($\beta\simeq1$) this frequency may acquire high
values. For instance, we have 
\begin{equation}
\beta\simeq1-\frac{\varepsilon_{0}^{2}}{2\varepsilon^{2}}\,\,,\,\,\omega\simeq2\omega_{0}\frac{\varepsilon}{\varepsilon_{0}}\,\,\,,\label{2}
\end{equation}
where $\varepsilon_{0}\simeq1GeV$ is the nucleon rest energy; for
$\varepsilon=1TeV$ we get a photon frequency $\omega\simeq2\times10^{3}\omega_{0}$
($\gamma=(1-\beta^{2})^{-1/2}\simeq\varepsilon/\varepsilon_{0}=10^{3}$).
We can see that for a $1eV$-laser we get $2keV$-photons in the rest
frame of the accelerated nucleus; for a $10keV$-free electron laser
we get $20MeV$-photons, etc. The effect is tunable by varying the
energy of the accelerated ions. This idea has been discussed in relation
to hydrogen-like accelerated heavy ions, which may scatter resonantly
$X$- or gamma-rays photons.$^{27}$ Similarly, a frequency up-shift
was discussed for photons reflected by a relativistically flying plasma
mirror generated by the laser-driven plasma wakefield,$^{28}$ or
photons in the rest frame of an ultra-relativistic electron beam.$^{24,29}$

For a typical laser radiation (see, for instance, ELI-NP project,$^{30}$)
we take a photon energy $\hbar\omega_{0}=1eV$ (wavelength $\lambda\simeq1\mu m$),
an energy $\mathcal{E}=50J$ and a pulse duration $\tau=50fs$. The
pulse length is $l=15\mu m$ (cca $15$ wavelengths), the power is
$P=10^{15}w$ ($1$ pettawatt). For a $d^{2}=(15\mu m)^{2}$-pulse
cross-sectional area the intensity is $I=P/d^{2}=4\times10^{20}w/cm^{2}$.
The electric field is $E\simeq10^{9}statvolt/cm$ ($1statvolt/cm=3\times10^{4}V/m$)
and the magnetic field is $H=10^{9}Gs$ ($1Ts=10^{4}Gs$). These are
very high fields (higher than atomic fields). The (ultra-) relativistic
ion will see a shortened pulse of length $l^{'}=\sqrt{1-\beta^{2}}l$,
with a shortened duration $\tau^{'}=\sqrt{1-\beta^{2}}\tau$ and an
energy $\mathcal{E}^{'}=\mathcal{E}\sqrt{(1+\beta/(1-\beta)}$ (the
number of photons $N_{ph}\simeq10^{20}$ is invariant). It follows
that the power and intensity are increased by the factor $(1-\beta)^{-1}$
($\simeq2\gamma^{2}$) and the fields are increased by the factor
$(1-\beta)^{-1/2}$; for instance, $E^{'}=E/\sqrt{1-\beta}=\sqrt{2}(\varepsilon/\varepsilon_{0})E\simeq10^{12}statvolt/cm$;
this figure is two orders of magnitude below Schwinger limit. 

A higher enhancement can be obtained by taking into account the aberration
of light, even from a collimated laser.$^{31-33}$ Indeed, for a cross-sectional
beam area $D^{2}=(0.5mm)^{2}$ we get an intensity $I=P/D^{2}=4\times10^{17}w/cm^{2}$
and an electric field $E\simeq5\times10^{7}statvolt/cm$ (all the
other parameters being the same). In the rest frame of the ion the
power increases by a factor $(1-\beta)^{-1}$, as before, but the
cross-sectional area $D^{'2}$ of the beam, decreases by a factor
$(1-\beta)/(1+\beta)$ ($\simeq1/4\gamma^{2}$), as a consequence
of the \textquotedbl{}forward beaming\textquotedbl{} (aberration of
light);$^{28}$ we have $D^{'2}=D^{2}(1-\beta)/(1+\beta)$, which
leads to an enhancement factor $(1+\beta)/(1-\beta)^{2}$ for intensity
and a factor $(1+\beta)^{1/2}/(1-\beta)$ ($\simeq2\sqrt{2}\gamma^{2}$)
for field. We get, for instance, $I^{'}\simeq3\times10^{24}w/cm^{2}$
and an electric field $E^{'}\simeq2\sqrt{2}\gamma^{2}E\simeq10^{14}statvolt/cm$. 

Similarly, we can take as typical parameters for a free electron laser
the photon energy $\hbar\omega_{0}=10keV$, the pulse duration $\tau=50fs$
and a much lower energy $\mathcal{E}=5\times10^{-5}J$ (power $P=10Gw$);
the fields may decrease by $3$ orders of magnitude, but still they
are very high $(10^{9}-10^{11}statvolt/cm$) in the rest frame of
the accelerated ion. 

Under these circumstances, the photons can attain energies sufficiently
high for photonuclear reactions, or giant dipole resonances, with
additional features arising from the electron-positron pair creation,
vacuum polarization, etc; indeed, above $\simeq1MeV$ the pair creation
in the Coulomb field of the atomic nucleus becomes important. Vacuum
polarization effects at very high intensity fields and high field
frequency are still insufficiently explored. Beside, all these happen
in two particular cirumstances: very short times and very high electromagnetic
fields. We discuss here the effect of these particular circumstances
on typical phenomena related to photon-nucleus interaction.

\section{\noindent Nuclear transitions }

Let us cosider an ensemble of interacting particles, some of them
with electric charge, like protons and neutrons in the atomic nucleus,
subjected to an external radiation field. We envisage quantum processes
driven by field energy quantum of the order $\hbar\Omega=10MeV$,
as discussed above. First, we note that the motion of the particles
at this energy is non-relativistic, since the particle rest energy
$\simeq1GeV$ is much higher than the energy quantum (we can check
that the acceleration $qE/m$ is much smaller than the \textquotedbl{}relativistic
acceleration\textquotedbl{} $c\Omega$, where $q$ and $m$ is the
particle charge and, respectively, mass and $E$ denotes he electric
field). Consequently, we start with the classical lagrangian $L=mv^{2}/2-V+q\mathbf{vA}/c-q\Phi$
of a particle with mass $m$ and charge $q$, moving in the potential
$V$ and subjected to the action of an electromagnetic field with
potentials $\Phi$ and $\mathbf{A}$; $\mathbf{v}$ is the particle
velocity. We get immediately the momentum $\mathbf{p}=m\mathbf{v}+q\mathbf{A}/c$
and the hamiltonian 
\begin{equation}
H=\frac{1}{2m}p^{2}+V-\frac{q}{mc}\mathbf{pA}+\frac{q^{2}}{2mc^{2}}A^{2}+q\Phi\,\,.\label{3}
\end{equation}
Usually, the particle hamiltonian $p^{2}/2m+V$ is separated and quantized
($V$ may be viewed as the mean-field potential of the nucleus), and
the remaining terms are treated as a perturbation. In the first order
of the perturbation theory we limit ourselves to the external radiation
field, which is considered sufficiently weak. Consequently, we put
$\mathbf{A}=\mathbf{A}_{0}$ and $\Phi=0$ in equation (\ref{3})
and take approximately $\mathbf{p}\simeq m\mathbf{v}$. We get the
well known interaction hamiltonian
\begin{equation}
H_{1}=-\frac{q}{c}\mathbf{v}\mathbf{A}_{0}=-\frac{1}{c}\mathbf{JA}_{0}\,\,\,,\label{4}
\end{equation}
 where $\mathbf{J}=q\mathbf{v}$ is the current; in the non-relativistic
limit we include also the spin currents in $\mathbf{J}$. If we leave
aside the spin currents, the interaction hamiltonian given by equation
(\ref{4}) can also be written as $q\mathbf{r}(d\mathbf{A}_{0}/dt)/c$.
Usually, the field does not depend on position over the spatial extension
of the ensemble of particles. Indeed, in the present case the wavelength
of the quantum $\hbar\Omega=10MeV$ is $\lambda\simeq10^{-12}cm$,
which is larger than the nucleus dimension $\simeq10^{-13}cm$; therefore
we may neglect the spatial variation of the field and write the interaction
hamiltonian as 
\begin{equation}
H_{1}=\frac{q}{c}\mathbf{r}\frac{d\mathbf{A}_{0}}{dt}=\frac{q}{c}\mathbf{r}\frac{\partial\mathbf{A}_{0}}{\partial t}=-q\mathbf{r}\mathbf{E}_{0}=-\mathbf{d}\mathbf{E}_{0}\,\,\,,\label{5}
\end{equation}
 where $\mathbf{d}=q\mathbf{r}$ is the dipole moment. This is the
well-known dipole approximation. For an ensemble of $N$ particles
we write the interaction hamiltonian given by equation (\ref{4})
as 
\begin{equation}
H_{1}=-\frac{1}{c}\sum_{i}\mathbf{J}_{i}\mathbf{A}_{0}\label{6}
\end{equation}
 (within the dipole approximation) and its matrix elements between
two states $a$ and $b$ are given by
\begin{equation}
\begin{array}{c}
H_{1}(a,b)=-\frac{1}{c}\mathbf{J}(a,b)\mathbf{A}_{0}=\\
\\
=-\frac{1}{c}[\sum_{i}\int d\mathbf{r}_{1}...d\mathbf{r}_{i}...d\mathbf{r}_{N}\psi_{a}^{*}(\mathbf{r}_{1}..\mathbf{r}_{i}..\mathbf{r}_{N})\mathbf{J}_{i}\psi_{b}(\mathbf{r}_{1}..\mathbf{r}_{i}..\mathbf{r}_{N})]\mathbf{A}_{0}\,\,\,,
\end{array}\label{7}
\end{equation}
 where $\psi_{a,b}$ are the wavefunctions of the two states $a$
and $b$; the notation $\mathbf{r}_{i}$ in equation (\ref{7}) includes
also the spin variable. As it is well known, the transition amplitude
is given by 
\begin{equation}
c_{ab}=-\frac{i}{\hbar}\int dtH_{1}(a,b)e^{i\omega_{ab}t}\,\,\,,\label{8}
\end{equation}
where $\omega_{ab}=(E_{a}-E_{b})/\hbar$ is the frequency associated
to the transition between the two states $a$ and $b$ with energies
$E_{a}$ and, respectively, $E_{b}.$ We take 
\begin{equation}
\mathbf{A}_{0}(t)=\mathbf{A}_{0}e^{-i\Omega t}+\mathbf{A}_{0}^{*}e^{i\Omega t}\label{9}
\end{equation}
 (with $\Omega>0$) and note that the pulse duration $\tau^{'}=\sqrt{1-\beta^{2}}\tau\simeq5\times10^{-17}s$
is much longer than the transition time $1/\Omega\simeq10^{-22}s$;
we can extend the integration in equation (\ref{8}) to infinity and
get 
\begin{equation}
c_{ab}=\frac{2\pi i}{\hbar c}\mathbf{J}(a,b)\mathbf{A}_{0}\delta(\omega_{ab}-\Omega)\,\,;\label{10}
\end{equation}
 making use of $\delta(\omega=0)=t/2\pi$, we get the number of transitions
per unit time 
\begin{equation}
P_{ab}=\left|c_{ab}\right|^{2}/t=2\pi\left|\frac{\mathbf{J}(a,b)\mathbf{A}_{0}}{\hbar c}\right|^{2}\delta(\omega_{ab}-\Omega)\,\,.\label{11}
\end{equation}
This is a standard calculation. Usually, the field and the wavefunctions
of the atomic nuclei are decomposed in electric and magnetic multiplets,
and the selection rules of conservation of the parity and the angular
momentum are made explicit (see, for instance,$^{34}$). It relates
to the absorption (emission) of one photon. 

It is worth estimating the number of transitions per unit time as
given by equation (\ref{11}). First, we may approximate $J(a,b)$
by $qv$. For an energy $\hbar\Omega=10MeV$ and a rest energy $1GeV$
we have $v/c=10^{-1}$. Next, from $\mathbf{E}_{0}=(-1/c)\partial\mathbf{A}_{0}/\partial t$
we deduce $A_{0}\simeq10^{-3}statvolt$ (for $E_{0}=10^{9}statvolt/cm$
and $\Omega=10^{22}s^{-1}$); it follows that the particle energy
in this field is $qA_{0}\simeq1eV$ (which is a very small energy).
We get from equation (\ref{11}) $P_{ab}\simeq(10^{28}/\Delta\Omega)s^{-1}$,
where $\Delta\Omega\simeq1/\tau^{'}\simeq10^{16}s^{-1}$ is the uncertainty
in the pulse frequency, such that the number of transitions per unit
time is $P_{ab}\simeq10^{12}s^{-1}$(much smaller than $\Omega=10^{22}s^{-1}$).
We can see that, under these circumstances, the first-order calculations
of the perturbation theory are justified. 

For higher fields we should include the second-order terms in the
interaction hamiltonian given by equation (\ref{3}); this second-order
interaction hamiltonian reads 
\begin{equation}
H_{2}=-\frac{q^{2}}{2mc^{2}}\mathbf{A}_{0}^{2}\,\,.\label{12}
\end{equation}
 We can see that within the dipole approximation this interaction
does not contribute to the transition amplitude, since the field does
not depend on position and the wavefunctions are orthogonal. For field
wavelengths shorter than the dimension of the ensemble of particles
(\emph{i.e.,} beyond the dipole approximation) we write 
\begin{equation}
\mathbf{A}_{0}(\mathbf{r},t)=\mathbf{A}_{0}e^{-i\Omega t+i\mathbf{kr}}+\mathbf{A}_{0}^{*}e^{i\Omega t-i\mathbf{kr}}\,\,\,,\label{13}
\end{equation}
 where $\mathbf{k}=\Omega/c$ is the wavevector, and get 
\begin{equation}
H_{2}(a,b)=-\frac{q^{2}}{2mc^{2}}\left[A_{0}^{2}(a,b)e^{-2i\Omega t}+A_{0}^{*2}(b,a)e^{2i\Omega t}\right]\,\,\,,\label{14}
\end{equation}
 where 
\begin{equation}
A_{0}^{2}(a,b)=[\sum_{i}\int d\mathbf{r}_{1}...d\mathbf{r}_{i}...d\mathbf{r}_{N}\psi_{a}^{*}(\mathbf{r}_{1}..\mathbf{r}_{i}..\mathbf{r}_{N})e^{2i\mathbf{k}\mathbf{r}_{i}}\psi_{b}(\mathbf{r}_{1}..\mathbf{r}_{i}..\mathbf{r}_{N})]A_{0}^{2}\,\,.\label{15}
\end{equation}
 This interaction gives rise to two-photon pocesses, with the transition
amplitude 
\begin{equation}
c_{ab}=\frac{2\pi i}{\hbar}\frac{q^{2}}{2mc^{2}}A_{0}^{2}(a,b)\delta(\omega_{ab}-2\Omega)\,\,.\label{16}
\end{equation}
 Comparing the transition amplitudes produced by the interaction hamiltonians
$H_{1}$ (equation (\ref{10})) and $H_{2}$ (equation (\ref{16}))
we may get an approximate criterion: $qA_{0}/mc^{2}$ (two-photons)
compared with $v/c$ (one photon). Since $v/c\simeq10^{-1}$ (as estimated
above), we should have $qA_{0}>10^{-1}\times1GeV=100MeV$ in order
to get a relevant contribution from two-photon processes. As estimated
above, $qA_{0}\simeq1eV$, so we can see that the second-order interaction
hamiltonian and the two-photon processes bring a very small contribution
to the transition amplitudes.

\section{\noindent Giant dipole resonance }

There is another process of excitation of the ensemble of particles
described by the hamiltonian given by equation (\ref{3}). Indeed,
let us write the interaction hamiltonian 
\begin{equation}
H_{int}=-\frac{q}{mc}\mathbf{pA}+\frac{q^{2}}{2mc^{2}}A^{2}+q\Phi\,\,\,,\label{17}
\end{equation}
 or 
\begin{equation}
H_{int}=-\frac{q}{c}\mathbf{vA}-\frac{q^{2}}{2mc^{2}}A^{2}+q\Phi\,\,.\label{18}
\end{equation}
 Under the action of the electromagnetic field the mobile charges
(\emph{e.g.,} protons in atomic nucleus) acquire a displacement $\mathbf{u}$,
which, in general, is a function $\mathbf{u}(\mathbf{r},t)$ of position
and time. This is a collective motion associated with the particle-density
degrees of freedom; in the limit of long wavelengths (\emph{i.e.}
for $\mathbf{u}$ independent of position) it is the motion of the
center of mass of the charges. Therefore, an additional velocity $\dot{\mathbf{u}}$
should be included in equation (\ref{18}). It is easy to see that
this $\mathbf{u}$-motion implies a variation $\rho_{p}=-nqdiv\mathbf{u}$
of the (volume) charge density and a current density $\mathbf{j}_{p}=nq\dot{\mathbf{u}}$,
where $n$ is the concentration of mobile charges. Obviously, these
are polarization charge and current densities (the suffix $p$ comes
from \textquotedbl{}polarization\textquotedbl{}). The charge and current
densities $\rho_{p}$ and $\mathbf{j}_{p}$ give rise to an internal,
polarization electromagnetic field, with the potentials $\mathbf{A}_{p}$
and $\Phi_{p}$ (related through the Lorenz gauge $div\mathbf{A}_{p}+(1/c)\partial\Phi_{p}/\partial t=0$),
which should be added to the potential of the external field in equation
(\ref{18}). Indeed, the retardation time $t_{r}=a/c\simeq10^{-23}s$,
where $a\simeq10^{-13}cm$ is the dimension of the atomic nucleus,
is shorter than the excitation time $\Omega^{-1}=10^{-22}s$, so the
atomic nucleus gets polarized. In particular the scalar potential
$\Phi$ in equation (\ref{18}) is the polarization scalar potential
$\Phi_{p}$. We get 
\begin{equation}
H_{int}=H_{1}-\frac{1}{c}\mathbf{J}\mathbf{A}_{p}-\frac{q}{c}\dot{\mathbf{u}}(\mathbf{A}_{0}+\mathbf{A}_{p})-\frac{q^{2}}{2mc^{2}}(\mathbf{A}_{0}+\mathbf{A}_{p})^{2}+q\Phi_{p}\,\,\,,\label{19}
\end{equation}
 where $H_{1}$ is given by equation (\ref{4}). Within the dipole
approximation we may take $\mathbf{u}$ independent of position, except
for the surface of the particle ensemble, where the density falls
abruptly to zero. A similar behaviour extends to the vector and scalar
polarization potentials (inside the ensemble); in addition, through
the Lorenz gauge, the scalar potential $\Phi_{p}$ can be taken independent
of time within this approximation. The surface effects can be neglected
as regards the scalar product of two orthogonal wavefunctions. All
these simplifications amount to neglecting all the terms in equation
(\ref{19}) except the first two; therefore, we are left with 
\begin{equation}
H_{int}\simeq H_{1}+H_{1p}\,\,,\,\, H_{1p}=-\frac{1}{c}\mathbf{J}\mathbf{A}_{p}\,\,;\label{20}
\end{equation}
 in order to get $\mathbf{A}_{p}$ we need a dynamics for the displacement
field $\mathbf{u}$. 

We can construct a dynamics for the displacement field $\mathbf{u}$
by assuming that it is subjected to internal forces of elastic type,
characterized by frequency $\omega_{c}$; the (non-relativistic) equation
of motion is given by 
\begin{equation}
m\ddot{\mathbf{u}}=q(\mathbf{E}_{0}+\mathbf{E}_{p})-m\omega_{c}^{2}\mathbf{u}\,\,\,,\label{21}
\end{equation}
 where $\mathbf{E}_{0}=-(1/c)\partial\mathbf{A}_{0}/\partial t$ is
the external electric field and $\mathbf{E}_{p}$ is the polarization
electric field. Within the dipole approximation, Gauss's equation
$div\mathbf{E}_{p}=4\pi\rho_{p}=-4\pi nqdiv\mathbf{u}$ gives $\mathbf{E}_{p}=-4\pi nq\mathbf{u}$
for matter of infinite extension (polarization $\mathbf{P}=nq\mathbf{u}$).
For polarizable bodies of finite size there appears a (de-) polarizing
factor $f$ within the same dipole approximation, as a consequence
of surface charges (for instance, $f=1/3$ for a sphere). Therefore,
we can write equation (\ref{21}) as 
\begin{equation}
\ddot{\mathbf{u}}+(\omega_{c}^{2}+f\omega_{p}^{2})\mathbf{u}=\frac{q}{m}\mathbf{E}_{0}\,\,\,,\label{22}
\end{equation}
where $\omega_{p}=\sqrt{4\pi nq^{2}/m}$ is the plasma frequency.
For nucleons we can estimate $\hbar\omega_{p}\simeq Z^{1/2}MeV$,
where $Z$ is the atomic number. An estimation for the characteristic
frequency $\omega_{c}$ can be obtained from $m\omega_{c}^{2}d^{2}/2=\mathcal{E}_{c}(d/a)$,
where $d$ is the displacement amplitude, $a$ is the dimension of
the nucleus and $\mathcal{E}_{c}$ ($\simeq7-8MeV$) is the mean cohesion
energy per nucleon; the maximum value of $d$ is the mean inter-particle
separation distance $d=a/A^{1/3}$, where $A$ is the mass number.
We get $\hbar\omega_{c}\simeq10A^{1/6}MeV$. It is convenient to introduce
the frequency $\Omega_{0}=(\omega_{c}^{2}+f\omega_{p}^{2})^{1/2}$,
which, as we can see from the preceding estimations, is of the order
of $10MeV$, and write the equation of motion (\ref{22}) as 
\begin{equation}
\ddot{\mathbf{u}}+\Omega_{0}^{2}\mathbf{u}=\frac{q}{m}\mathbf{E}_{0}\,\,.\label{23}
\end{equation}

This is the equation of motion of a linear harmonic oscillator under
the action of an external force $q\mathbf{E}_{0}$. Making use of
equation (\ref{9}), we get the external field 
\begin{equation}
\mathbf{E}_{0}=\frac{i\Omega}{c}\mathbf{A}_{0}e^{-i\Omega t}-\frac{i\Omega}{c}\mathbf{A}_{0}^{*}e^{i\Omega t}\,\,;\label{24}
\end{equation}
 for frequency $\Omega$ approaching the oscillator frequency $\Omega_{0}$
the motion described by equation (\ref{23}) is a classical motion,
and we get 
\begin{equation}
\mathbf{u}=-\frac{iq\Omega}{mc}\cdot\frac{1}{\Omega^{2}-\Omega_{0}^{2}}\left(\mathbf{A}_{0}e^{-i\Omega t}-\mathbf{A}_{0}^{*}e^{i\Omega t}\right)\,\,.\label{25}
\end{equation}
 According to the discussion made above, the polarization field is
\begin{equation}
\mathbf{E}_{p}=-4\pi fnq\mathbf{u}=\frac{if\omega_{p}^{2}\Omega}{c}\cdot\frac{1}{\Omega^{2}-\Omega_{0}^{2}}\left(\mathbf{A}_{0}e^{-i\Omega t}-\mathbf{A}_{0}^{*}e^{i\Omega t}\right)\label{26}
\end{equation}
 and the corresponding vector potential is 
\begin{equation}
\mathbf{A}_{p}=\frac{f\omega_{p}^{2}}{\Omega^{2}-\Omega_{0}^{2}}\left(\mathbf{A}_{0}e^{-i\Omega t}+\mathbf{A}_{0}^{*}e^{i\Omega t}\right)\,\,.\label{27}
\end{equation}

A damping factor $\Gamma$ can be included in equation (\ref{23}),
\begin{equation}
\ddot{\mathbf{u}}+\Omega_{0}^{2}\mathbf{u}+\Gamma\dot{\mathbf{u}}=\frac{q}{m}\mathbf{E}_{0}\,\,\,,\label{28}
\end{equation}
 and we can write the solution as 
\begin{equation}
\mathbf{u}=-\frac{q}{m}\mathbf{E}_{0}\frac{1}{\Omega^{2}-\Omega_{0}^{2}+i\Omega\Gamma}e^{-i\Omega t}+c.c.\,\,;\label{29}
\end{equation}
the polarization reads 
\begin{equation}
\mathbf{P}=nqf\mathbf{u}=-\frac{f\omega_{p}^{2}}{4\pi}\frac{1}{\Omega^{2}-\Omega_{0}^{2}+i\Omega\Gamma}\mathbf{E}_{0}e^{-i\Omega t}+c.c.\,\,\,,\label{30}
\end{equation}
 so that we can define the polarizability 
\begin{equation}
\alpha=-\frac{f\omega_{p}^{2}}{4\pi}\frac{1}{\Omega^{2}-\Omega_{0}^{2}+i\Omega\Gamma}\,\,.\label{31}
\end{equation}
Therefore, the vector potential $\mathbf{A}_{p}$ given by equation
(\ref{27}) can be written as 
\begin{equation}
\mathbf{A}_{p}=-4\pi\left(\alpha\mathbf{A}_{0}e^{-i\Omega t}+\alpha^{*}\mathbf{A}_{0}^{*}e^{i\Omega t}\right)\,\,.\label{32}
\end{equation}

Now, we can estimate the transition amplitude between two states $a$
and $b$, making use of the interaction hamiltonian $H_{1p}$ given
by equation (\ref{20}). We get the amplitude 
\begin{equation}
c_{ab}=-\frac{8\pi^{2}i}{\hbar c}\alpha\mathbf{J}(a,b)\mathbf{A}_{0}\delta(\omega_{ab}-\Omega)\label{33}
\end{equation}
 and the number of transitions per unit time 
\begin{equation}
P_{ab}=32\pi^{3}\left|\frac{\mathbf{J}(a,b)\mathbf{A}_{0}}{\hbar c}\right|^{2}\left|\alpha\right|^{2}\delta(\omega_{ab}-\Omega)\,\,.\label{34}
\end{equation}
Comparing this result with equation (\ref{11}) we can see that, apart
from a numerical factor, the rate of polarization-driven transitions
are modified by the factor
\begin{equation}
\left|\alpha\right|^{2}=\left(\frac{f\omega_{p}^{2}}{4\pi}\right)^{2}\frac{1}{(\Omega^{2}-\Omega_{0}^{2})^{2}+\Omega^{2}\gamma^{2}}\,\,.\label{35}
\end{equation}
This is a typical resonance factor, which indicates that the polarization
of the particle ensemble is important for $\Omega\simeq\Omega_{0}$
(at resonance), where the ensemble can be disrupted. Obviously, this
is a giant dipole resonance.$^{35,36}$ For $\Omega$ far away from
the resonance frequency $\Omega_{p}$ the polarization is practically
irrelevant, and it may be neglected in comparison with the transitions
brought about by the interaction hamiltonian $H_{1}$ (equation (\ref{11})).
It is worth noting that we can define an electric susceptibility $\chi$
and a dielectric function $\varepsilon$ for the polarizable ensemble
of particles, by combining equations (\ref{4}), (\ref{20}) and (\ref{32}).
We get 
\begin{equation}
H_{1}+H_{1p}=-\frac{1}{c}\mathbf{J}\left[(1-4\pi\alpha)\mathbf{A}_{0}e^{-i\Omega t}+c.c.\right]=-\frac{1}{c}\mathbf{J}\left[\frac{1}{\varepsilon}\mathbf{A}_{0}e^{-i\Omega t}+c.c\right]\,\,\,,\label{36}
\end{equation}
 since $1-4\pi\alpha=(1+4\pi\chi)^{-1}=1/\varepsilon$, as expected
(according to their definitions, we have $\mathbf{P}=\alpha\mathbf{E}_{0}=\chi(\mathbf{E}_{0}-4\pi\mathbf{P})$,
where $\mathbf{P}$ is the polarization,\emph{ i.e.} the dipole moment
per unit volume). Therefore, the total interaction hamiltonian is
proportional to $1/\varepsilon=(\Omega^{2}-\omega_{c}^{2})/(\Omega^{2}-\Omega_{0}^{2})$,
and we note that, beside the $\Omega_{0}$-pole, it has a zero for
$\Omega=\omega_{c}$, where the transitions are absent. 

A similar description holds for ions (or neutral atoms) in an external
electromagnetic field. Perhaps the most interesting case is a neutral,
heavy atom, for which we can estimate the plasma energy $\hbar\omega_{p}\simeq10Z^{1/2}eV$.
For the cohesion energy per electron we can use the Thomas-Fermi estimation
$16Z^{7/3}/ZeV=16Z^{4/4}eV$, which leads to $\hbar\omega_{c}\simeq13Z^{5/6}eV$.
We can see that the typical scale energy where we may expect to occur
a giant dipole resonance is $\hbar\Omega_{0}\simeq1keV$. However,
the motion of the electrons under the action of a high-intensity electromagnetic
field is relativistic (see, for instance,$^{37}$).

\section{\noindent Discussion and conclusions }

The direct photon-nucleus coupling processes described here are hampered
by electron-positron pairs creation in the Coulomb field of the nucleus.
For photons of energy $\hbar\Omega=10MeV$ we may consider the (ultra-)
relativistic limit of the pair creation cross-section. As it is well
known,$^{38,39}$ in this case the cross-section is derived within
the Born approximation, the pair partners are generated mainly in
the forward direction, they have not very different energies from
one another and the recoil momentum (energy) trasmitted to the nucleus
is small. For bare nuclei (absence of screening) the total cross-section
of pair production is given by
\begin{equation}
\sigma_{pair}=\frac{Z^{2}r_{0}^{2}}{137}\left(\frac{28}{9}\ln\frac{2\hbar\Omega}{mc^{2}}-\frac{218}{27}\right)\simeq10^{-28}Z^{2}cm^{2}\,\,\,,\label{37}
\end{equation}
 where $r_{0}=e^{2}/mc^{2}$ is the classical electron radius, $-e$
is the electron charge and $m$ is the electron mass. We can get an
order of magnitude estimation of the efficiency of the processes described
here by comparing this cross-section with the nuclear cross-section
$a^{2}$$\simeq10^{-26}cm^{2}$. We can see that $\sigma_{pair}/a^{2}\simeq10^{-2}Z^{2}$,
which may go as high as $10^{2}$ for heavy nuclei. 

In conclusion, we may say that in the rest frame of (ultra-) relativistically
accelerated heavy ions (atomic nuclei) the electromagnetic radiation
field produced by high-power optical or free electron lasers may acquire
high intensity and high energy, suitable for photonuclear reactions.
In particular, the excitation of dipole giant resonance may be achieved.
Nuclear transitions are analyzed here under such particular circumstances,
including both one- and two-photon processes. It is shown that the
perturbation theory is applicable, although the field intensity is
high, since the interaction energy is low (as a consequence of the
high frequency) and the interaction time (pulse duration is short).
It is also shown that the giant nuclear dipole resonance is driven
by the nuclear (electrical) polarization degrees of freedom, whose
dynamics may lead to disruption of the atomic nucleus when resonance
conditions are met. The concept of nuclear (electrical) polarization
is introduced, as well as the concept of nuclear electrical polarizability
and dielectric function. \\

\textbf{ACKNOWLEDGMENTS}

The authors are indebted to the members of the Seminar of the Laboratory
of Theoretical Physics at Magurele-Bucharest for useful discussions.
The collaborative atmosphere of the Institute for Physics and Nuclear
Engineering and the Institute for Lasers, Plasma and Radiation at
Magurele-Bucharest is also gratefully acknowledged. This work has
been supported by UEFISCDI Grants CORE Program \#09370102/2009 and
PN-II-ID-PCE-2011-3-0958 of the Romanian Governmental Agency of Scientfic
Research. 

Both authors contributed equally to this work.\\

$^{1}$M. Beard, S. Frauendorf, B. Kampfer, R. Schwengner, and M.
Wiescher, \textquotedbl{}Photonuclear and radiative-capture reaction
rates for nuclear astrophysics and transmutation: $^{92-100}Mo$,
$^{88}Sr$, $^{90}Zr$, and $^{139}La$,\textquotedbl{} Phys. Rev.
\textbf{C85}, 065108 (2012).

$^{2}$T. D. Thiep, T. T. An, N. T. Khai, P. V. Cuong, N. T. Vinh,
A. G. Belov, and O. D. Maslo, \textquotedbl{}Study of the isomeric
ratios in photonuclear reactions of natural Selenium induced by bremsstrahlungs
with end-point energies in the giant dipole resonance region,'' J.
Radioanal. Nucl. Chemistry \textbf{292}, 1035 (2012). 

$^{3}$A. Giulietti, N. Bourgeois, T. Ceccotti, X. Davoine, S. Dobosz,
P. D\textquoteright{}Oliveira, M. Galimberti, J. Galy, A. Gamucci,
D. Giulietti, L. A. Gizzi, D. J. Hamilton, E. Lefebvre, L. Labate,
J. R. Marques, P. Monot, H. Popescu, F. Reau, G. Sarri, P. Tomassini,
and P. Martin, \textquotedbl{}Intense $\gamma$-ray source in the
giant-dipole-resonance range driven by $10-Tw$ laser pulses,\textquotedbl{}
Phys. Rev. Lett. \textbf{101}, 105002 (2008).

$^{4}$V. G. Neudatchin, V. I. Kukulin, V. N. Pomerantsev, and A.
A. Sakharuk, \textquotedbl{}Generalized potential-model description
of mutual scattering of the lightest $p+d$, $d+^{3}He$ nuclei and
the corresponding photonuclear reactions,\textquotedbl{} Phys. Rev.
\textbf{C45}, 1512 (1992).

$^{5}$V. N. Litvinenko, B. Burnham, M. Emamian, N. Hower, J. M. J.
Madey, P. Morcombe, P. G. O'Shea, S. H. Park, R. Sachtschale, K. D.
Straub, G. Swift, P. Wang, Y. Wu, R. S. Canon, C. R. Howell, N. R.
Roberson, E. C. Schreiber, M. Spraker, W. Tornow, H. R. Weller, I.
V. Pinayev, N. G. Gavrilov, M. G. Fedotov, G. N. Kulipanov, G. Y.
Kurkin, S. F. Mikhailov, V. M. Popik, A. N. Skrinsky, N. A. Vinokurov,
B. E. Norum, A. Lumpkin, and B. Yang, \textquotedbl{}Gamma-Ray Production
in a Storage Ring Free-Electron Laser,\textquotedbl{} Phys. Rev. Lett.
\textbf{78}, 4569 (1997). 

$^{6}$S. Amano, K. Horikawa, K. Ishihara, S. Miyamoto, T. Hayakawa,
T. Shizuma, and T. Mochizuki, \textquotedbl{}Several-$MeV$ \textgreek{g}-ray
generation at new SUBARU by laser Compton backscattering,\textquotedbl{}
Nucl. Instr. Method \textbf{A602}, 337 (2009).

$^{7}$S. V. Bulanov, T. Zh Esirkepov, Y. Hayashi, M. Kando, H. Kiriyama,
J. K. Koga, K. Kondo, H. Kotaki, A. S. Pirozhhkov, S. S. Bulanov,
A. G. Zhidkov, P. Chen, D. Neely, Y. Kato, N. B. Narozhny, and G.
Korn, \textquotedbl{}On the design of experiments for the study of
extreme field limits in the interaction of laser with ultrarelativistic
electron beam,\textquotedbl{} Nucl. Instr. Meth. Phys. Res. \textbf{A660},
31 (2011).

$^{8}$C. Maroli, V. Petrillo, P. Tomassini, and L. Serafin, \textquotedbl{}Nonlinear
effects in Thomson backscattering,\textquotedbl{} Phys. Rev. Accel.
Beams \textbf{16}, 030706 (2013).

$^{9}$E. V. Abakumova, M. N. Achasov, D. E. Berkaev, V. V. Kaminsky,
N. Yu. Muchnoi, E. A. Perevedentsev, E. E. Pyata, and Yu. M. Shatunov,
\textquotedbl{}Backscattering of Laser Radiation on Ultrarelativistic
Electrons in a Transverse Magnetic Field: Evidence of $MeV$-Scale
Photon Interference,\textquotedbl{} Phys. Rev. Lett. \textbf{110},
140402 (2013).

$^{10}$S. S. Bulanov, C. B. Schroeder, E. Esarey, and W. P. Leemans,
\textquotedbl{}Electromagnetic cascade in high-energy electron, positron,
and photon interactions with intense laser pulses,\textquotedbl{}
Phys. Rev. \textbf{A87}, 062110 (2013).

$^{11}$K. Krajewska, C. Muller, and J. Z. Kaminski, \textquotedbl{}Bethe-Heitler
pair production in ultrastrong short laser pulses,\textquotedbl{}
Phys. Rev. \textbf{A87}, 062107 (2013).

$^{12}$S. Cipiccia, S. M. Wiggins, R. P. Shanks, M. R. Islam, G.
Vieux, R. C. Issac, E. Brunetti, B. Ersfeld, G. H. Welsh, M. P. Anania,
D. Maneuski, N. R. C. Lemos, R. A. Bendoyro, P. P. Rajeev, P. Foster,
N. Bourgeois, T. P. A. Ibbotson, P. A. Walker, V. O. Shea, J. M. Dias,
and D. A. Jaroszynski, \textquotedbl{}A tuneable ultra-compact high-power,
ultra-short pulsed, bright gamma-ray source based on bremsstrahlung
radiation from laser-plasma accelerated electrons,\textquotedbl{}
J. Appl. Phys. \textbf{111}, 063302 (2012).

$^{13}$A. Makinaga, K. Kato, T. Kamiyama, and K. Yamamoto, \textquotedbl{}Development
of a new bremsstrahlung source for nuclear astrophysics\textquotedbl{},
in \emph{The $10$th International Symposium on Origin of Matter and
Evolution of Galaxies, OMEG-2010}, Osaka,\emph{ }Japan, 8-10 March
2010\emph{,} edited by I. Tanihara, H. J. Ong, A. Tamii, T. Kishimoto,
S. Kubano, and T. Shima (AIP Conf. Proc. \textbf{1269}, 2010) pp.
394-396. 

$^{14}$S. Matinyan, \textquotedbl{}Lasers as a bridge between atomic
and nuclear physics,\textquotedbl{} Phys. Reps. \textbf{298}, 199
(1998).

$^{15}$K. W. D. Ledingham, P. McKenna, and R. P. Singhal, \textquotedbl{}Applications
for Nuclear Phenomena Generated by Ultra-Intense Lasers,\textquotedbl{}
Science \textbf{300}, 1107 (2003).

$^{16}$K. V. D. Ledingham and W. Galster, \textquotedbl{}Laser-driven
particle and photon beams and some applications,\textquotedbl{} New
J. Phys. \textbf{12},\textbf{ }045005 (2010).

$^{17}$W. P. Leemans, B. Nagler, A. J. Gonsalves, Cs. Toth, K. Nakamura,
C. G. R. Geddes, E. Esarey, C. B. Schroeder, and S. M. Hooker, \textquotedbl{}Particle
physics GeV electron beams from a centimetre-scale accelerator,\textquotedbl{}
Nature Physics \textbf{2}, 696 (2006).

$^{18}$M. Apostol and M. Ganciu, \textquotedbl{}Polaritonic pulse
and coherent $X$- and gamma rays from Compton (Thomson) backscattering,\textquotedbl{}
J. Appl. Phys. \textbf{109}, 013307 (2011).

$^{19}$G. A. Mourou, N. J. Fisch, V. M. Malkin, Z. Toroker, E. A.
Khazanov, A. M. Sergeev, T. Tajima, and T.B. Le Garrec, \textquotedbl{}Exawatt-Zettawatt
pulse generation and applications,\textquotedbl{} Optics Commun. \textbf{285},
720 (2012).

$^{20}$H. Schwoerer, J. Magill, and B. Beleites, eds., \emph{Lasers
and Nuclei: Applications of Ultrahigh Intensity Lasers in Nuclear
Science,} Springer Lectures Notes in Phys. \textbf{694} (Springer,
Berlin, Heidelberg, 2006).

$^{21}$C. A. Bertulani, S. R. Klein, and J. Nystrand, \textquotedbl{}Physics
of ultra-peripheral nuclear collisions,\textquotedbl{} Annual Rev.
Nucl. Particle Sci. \textbf{55}, 271 (2005).

$^{22}$A. Di Piazza, K. Z. Hatsagortsyan, and C. H. Keitel, \textquotedbl{}Nonperturbative
Vacuum-Polarization Effects in Proton-Laser Collisions,\textquotedbl{}
Phys. Rev. Lett. \textbf{100}, 010403 (2008). 

$^{23}$D. d\textquoteright{}Enterria and G. G. da Silveira, \textquotedbl{}Observing
Light-by-Light Scattering at the Large Hadron Collider,\textquotedbl{}
Phys. Rev. Lett. \textbf{111}, 080405 (2013).

$^{24}$D. L. Burke, R. C. Field, G. Horton-Smith, J. E. Spencer,
D. Walz, S. C. Berridge, W. M. Bugg, K. Shmakov, A. W. Weidemann,
C. Bula, K. T. McDonald, E. J. Prebys, C. Bamber, S. J. Boege, T.
Koffas, T. Kotseroglou, A. C. Melissinos, D. D. Meyerhofer, D. A.
Reis, and W. Ragg, \textquotedbl{}Positron Production in Multiphoton
Light-by-Light Scattering,\textquotedbl{} Phys. Rev. Lett. \textbf{79},
1626 (1997).

$^{25}$C. Muller, \textquotedbl{}Non-linear Bethe-Heitler pair creation
with attosecond laser pulses at the LHC,\textquotedbl{} Phys. Lett.
\textbf{B672}, 56 (2009). 

$^{26}$Large Hadron Collider, CERN, Geneva, http://cds.cern.ch/record/\\
1272417/files/ATL-GEN-SLIDE-2010-139.pdf

$^{27}$E. G. Bessonov and K.-J. Kim, \textquotedbl{}Gamma ray sources
based on resonant backscattering of laser beams with relativistic
heavy ion beams,\textquotedbl{} in \emph{Proc of the $16th$ Biennial
Particle Accelerator Conference}, Dallas, May 1995, vols. \textbf{1-5}
(1996) pp. 2895-2897.

$^{28}$S. V. Bulanov, T. Esirkepov, and T. Tajima, \textquotedbl{}Light
Intensification towards the Schwinger Limit,\textquotedbl{} Phys.
Rev. Lett. \textbf{91}, 085001 (2003). 

$^{29}$C. Bula, K. T. McDonald, E. J. Prebys, C. Bamber, S. Boege,
T. Kotseroglou, A. C. Melissinos, D. D. Meyerhofer, W. Ragg, D. L.
Burke, R. C. Field, G. Horton-Smith, A. C. Odian, J. E. Spencer, D.
Walz, S. C. Berridge, W. M. Bugg, K. Shmakov, and A. W. Weidemann,
\textquotedbl{}Observation of Nonlinear Effects in Compton Scattering,\textquotedbl{}
Phys. Rev. Lett. \textbf{76}, 3116 (1996).

\noindent $^{30}$Extreme Light Infrastructure-Nuclear Physics Project
(ELI-NP), \\
http://www.eli-np.ro/documents/ ELI-NP-WhiteBook.pdf, \\
http://www.extreme-light-infrastructure.eu.

$^{31}$A. Lampa, \textquotedbl{}Wie erscheint nach der Relativitatstheorie
ein bewegter Stab einem ruhenden Beobachter,\textquotedbl{} Z. Phys.
\textbf{27}, 138 (1924).

$^{32}$J. Terrell, \textquotedbl{}Invisibility of the Lorentz contraction,\textquotedbl{}
Phys. Rev. \textbf{116}, 1041 (1959).

$^{33}$R. Penrose, \textquotedbl{}The apparent shape of a relativistically
moving sphere,\textquotedbl{} Math. Proc. Cambridge Phil. Soc. \textbf{55},
137 (1959).

$^{34}$J. M. Blatt and V . F. Weisskopf, \emph{Theoretical Nuclear
Physics} (Dover, NY, 1979).

$^{35}$M. Goldhaber and E. Teller, \textquotedbl{}On nuclear dipole
vibrations,\textquotedbl{} Phys. Rev. \textbf{74}, 1046 (1948).

$^{36}$H. A. Weidenmuller, \textquotedbl{}Nuclear Excitation by a
Zeptosecond Multi-$MeV$ Laser Pulse,\textquotedbl{} Phys. Rev. Lett.
\textbf{106}, 122502 (2011). 

$^{37}$A. Di Piazza, C. Muller, K. Z. Hatsagortsyan, and C. H. Keitel,
\textquotedbl{}Extremely high-intensity laser interactions with fundamental
quantum systems,\textquotedbl{} Revs. Mod. Phys. \textbf{84}, 1177
(2012). 

$^{38}$W. Heitler, \emph{The Quantum Theory of Radiation} (Dover,
NY, 1984).

$^{39}$V. B. Berestetskii, E. M. Lifshitz, and L. P. Pitaevskii,
\emph{Quantum Electrodynamics}: L. Landau and E. Lifshitz, \emph{Course
of Theoretical Physics}, vol. 4 ( (Butterworth-Heinemann, Oxford,
1982).xt
\end{document}